# Entropy Generation by a Maxwell Demon in the Sequential Sorting of the Particles in an Ideal Gas

**Roger D. Jones[a,b], Sven G. Redsun[a], Roger E. Frye[a]**

[a]Complexica/CommodiCast
125 Lincoln Avenue, Suite 400
Santa Fe, NM 87501 USA
Roger.Jones@Complexica.com
www.complexica.com
www.commodicast.com

[b]Applied Theoretical Physics Division
Los Alamos National Laboratory
Los Alamos, NM 87545 USA

**October 19, 2003**

## Abstract

**This paper revisits the Maxwell Demon Problem. Representing the demon with a simple physical computer composed of a single memory element, we demonstrate that the average minimum entropy increase of the universe due to sorting of particles with a Maxwell Demon is $\eta \approx 0.8400$ bits/particle for particles that are initially randomly distributed.**

## 1. Introduction

This paper examines a very old problem in a modern information-age context. The question originated with Maxwell nearly 140 years ago and can be stated simply as "Can measurement and intelligence invalidate the Second Law of Thermodynamics?" In Maxwell's original problem he imagined an intelligent demon that could separate a box of particles into fast particles and slow particles, the fast particles segregated to one side of the box and the slow to the other. Here, we perform an entropy accounting of a simplified version of Maxwell's original problem and show that the minimum entropy increase per particle of the universe due to sorting of particles by a simple physical computer is given by

$$\eta \approx 0.8400 \text{ bits/particle} \tag{1.1}$$

for particles that are initially randomly distributed throughout the box. We show elsewhere[1] that a closed form solution exists for Eq. (1.1) given by

$$\eta = \frac{\pi^2}{12\ln(2)} - \frac{\ln(2)}{2}. \tag{1.2}$$

## 2. A Simplified Version of the Original Maxwell Demon

Maxwell[2][3][4] created his Demon in 1867 to help clarify the issues associated with the Second Law of Thermodynamics. In particular, Maxwell wished to address the question of the role of intelligence in the flow of entropy. The Demon was an intelligent microscopic creature that sat at a trapdoor separating a box into two sides. Particles inhabited both sides of the box. The Demon observed the particles and allowed fast particles to enter into one side of the box and slow particles to enter into the other side of the box. The entropy of the particles was thus decreased and a temperature gradient, capable of producing useful work, was created. The intelligent Demon seemed to violate the Second Law. Either the Second Law had to be abandoned or the entropy of the Demon had to increase to compensate for the decrease in entropy of the particles.

In this paper, we re-examine the Demon problem in a modern context (although we ignore quantum and relativistic effects). Our Demon is a simple physical computer. We simplify the problem somewhat from the original. Rather than fast and slow particles, we consider here particles labeled $A$ and $B$. These can be particles of slightly different chemical makeup or perhaps two slightly different isotopes. Therefore, we imagine that we have a box of $N$ particles identical in every way except that half of the particles are labeled $A$ and half are labeled $B$, There is a removable trapdoor that divides the box in half that, when closed, allows particles to access only the right, $R$, or left, $L$, side of the box and, when open, allows the particles to access the entire box. The trapdoor can be removed or installed in zero time and it absorbs no energy in the process. Initially, the particles are randomly scattered and evenly distributed throughout the box. The Demon's goal is to sort the particles so that all $A$ particles are on the left and all $B$ particles are on the right. The Demon looks at the next particle to intersect the trapdoor and determines if that particle is of type $A$ or $B$ and whether the particle is on side $R$ or $L$. As the Maxwell Demon observes the particles in the box he records relevant information about the particle onto a simple Memory Unit. The Demon then sends a signal to a Controller to either have the trapdoor closed, $c$, or open, $o$, in order to maintain the particle on its correct side or to let the particle pass to its correct side. We assume there is no energy or entropy flow associated with the particle measurement or in sending the signal to the trapdoor. There is energy and entropy flow associated with the writing and erasing of information onto the Memory Unit. We assume that after a particle passes through the trapdoor or is reflected by the trapdoor that it is mixed with the other particles on the same side of the box infinitely fast so that it has the same probability of intersecting the trapdoor as any other particle of its type on the same side of the box. We measure time by counting the number of particles that approach the trapdoor. Time $t$, therefore, is the time associated with the $t^{th}$ particle that approached the trapdoor.

## *3. Important Principles*

There are three important principles that have emerged in modern treatments of the Maxwell Demon problem. Any treatment of the Maxwell Demon problem must be consistent with these principles.

## Principle 1: Probabilistic and Deterministic Equivalence

The first principle, due to Zurek[5], states that an entropy calculation based on a traditional average over an ensemble of states yields the same results as a deterministic calculation of the number of bits required to describe the system. In applications it is often convenient to use a hybrid approach that utilizes both probabilistic and deterministic components. Zurek defines *physical entropy* as the sum of missing information determined from a probabilistic approach and the length, in bits, of the most concise record expressing the information already measured about the system (*algorithmic complexity* or *algorithmic entropy*).

## Principle 2: Information Destruction and Heat Generation

The second principle, due to Landauer[6], states that logically irreversible erasure of computer tape generates an minimum amount of heat approximately equal to $k_B T$ per bit. Landauer argued that to each logical state there corresponds a physical state. Logical irreversibility implies a reduction in physical degrees of freedom that results in dissipation. The original association of information and heat flow was made by Szilard[7]

## Principle 3: Reversible Logic and a Thermodynamically Reversible Computer

The third principle, due to Bennett[8], states that logically reversible computation can be thermodynamically reversible. Thus one can, in principle, write to an input tape, perform reversible computation while keeping an historical record, create a copy of the output, then reverse the computation reversibly erasing all tapes except for the copy of the output tape. As long as the output tape is not erased there is no heat transfer to the environment. Since all information for generation of the output tape has been destroyed, erasure of the output tape is logically irreversible. From Landauer's Principle (Principle 2) then, heat is transferred to the environment upon erasure of the copy of the output tape.

In this paper, we use all three principles. We use Zurek's Principle (Principle 1) to justify the use of a statistical ensemble approach for both the physical system and the computer. Bennett's Principle (Principle 3) is used to justify the use of a single output unit as the sole computer in our treatment. Finally, our result (Eq. (1.1)) is a specific instance of Landauer's Principle (Principle 2).

There is another principle that is important in this paper. The Maxwell Demon is part of a control loop. The Demon is not a dispassionate observer. He plays a crucial role in ordering the physical system. Quantitative entropy and energy accounting can only be performed if the complete feedback loop is considered. This feedback has been minimized in some previous treatments. It plays a central role in this paper.

## *4. The Physical System and Its Control*

The physical system with its control apparatus is illustrated in FIG 1. There are two actual physical entities, the system itself consisting of the box of particles and a partition in the middle of the box, and a computer, which we refer to here as a "Memory Unit." We identify the Memory Unit with the Maxwell Demon. There are two communication channels between the two physical entities, a measurement channel that communicates the state of a particle approaching the trapdoor to the computer (Memory Unit or Demon), and a control signal that communicates the desired state of the trapdoor to the system. Measurements are made on each particle as it approaches the trapdoor. The appropriate input signal is written to the Memory Unit, which sends the control signal to the trapdoor. The Memory Unit is immersed in a heat bath. The heat bath receives the excess entropy generated in the sorting process. We assume the system itself is thermally insulated and only interacts with its environment through the Memory Unit. We neglect any work or entropy generation associated with opening and closing the trapdoor, with measuring the state of the particles, communicating the measurements or the control signal, and with any computation other than writing and erasing information to the Memory Unit. We do, however, take into account the work and entropy generation associated with writing and erasing the Memory Unit and entropy/work associated with the particles.

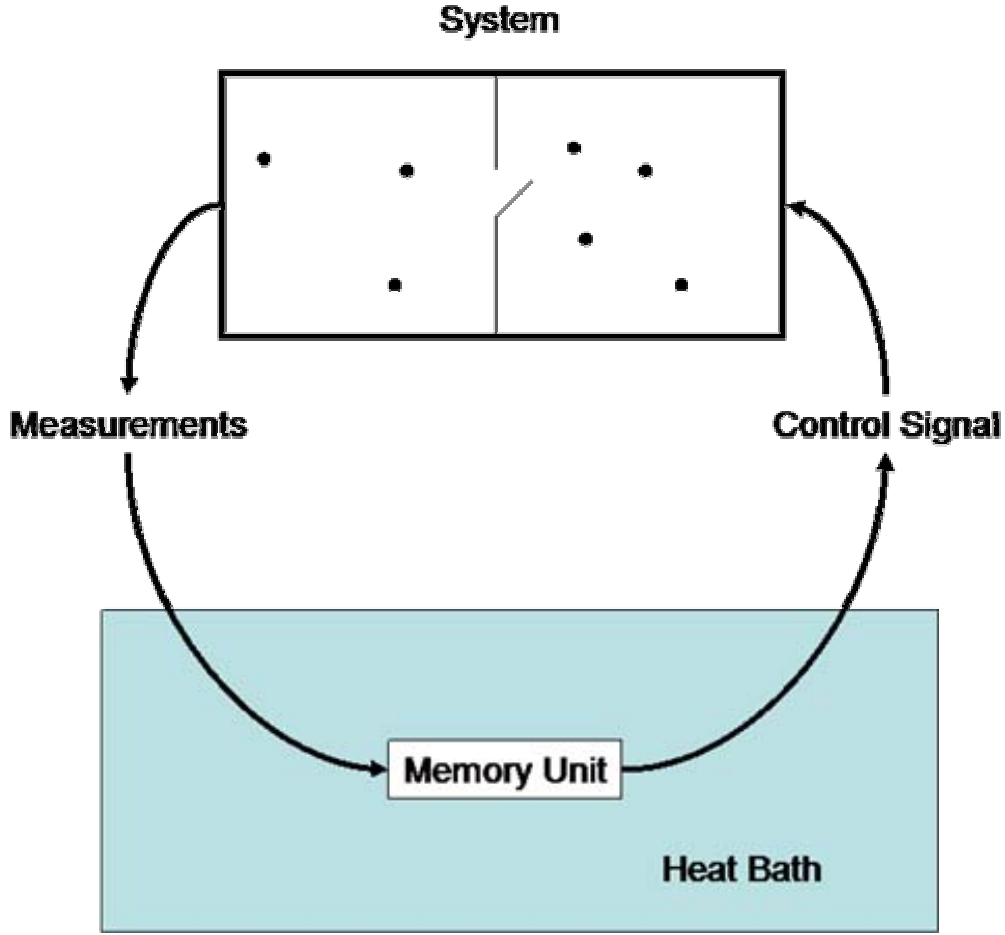

**FIG 1: Measurements are made on the t<sup>th</sup> particle approaching the trapdoor. The minimal information needed to control the trapdoor is written to a Memory Unit. The Memory Unit sends the appropriate control signal to the trapdoor. The trapdoor either opens or closes allowing the particle to pass or be reflected. The Memory Unit is then erased in preparation for the (t+1)<sup>th</sup> measurement. There is no dissipation in the measurement or control processes. There is irreversible heat flow to the bath upon erasure of the Memory Unit.**

Our goal is to design a computer or Memory Unit that orders the system in the most efficient manner, and to measure the entropy increase of the universe (represented here by the heat bath) that results from the sorting process.

We can use the theory of communication[9] to inform our design of the Memory Unit. The communication channels must have the capacity to transmit sufficient information to control the trapdoor. This will suggest the minimum amount of heat that must flow to the bath. To see this note that if at any given time $t$, the probability that the control signal will be open ($o$) is designated by $\Pi_t$, then the communication line that transmits the control signal must be able to transmit

$$I_t = -\Pi_t \lg(\Pi_t) - (1-\Pi_t)\lg(1-\Pi_t) \tag{4.1}$$

bits of information at time $t$. Here $\lg$ is the logarithm to the base 2. Equation (4.1) is the amount of information needed to determine whether the trapdoor should be open or closed. Since the control signal is the output of the Memory Unit, then, by Landauer's Principle, erasure of the output in preparation for the next measurement should release an amount of heat to the bath in proportion to the number of bits in the output, which in this case is just $I_t$. We thus expect the increase in entropy of the heat bath to increase by an amount approximately equal to $I_t$ (measured in bits).

We can carry these arguments a step further and place constraints on the optimal information capacity of the measurement communication channel. The measurements are the input to the Memory Unit. From Bennett's Principle we know that the inputs can contain information in excess of information in the output with no extra heat flow to the bath. The Memory Unit must, however, separate the useful information from the unnecessary information. This increases the *algorithmic complexity* (the smallest number of bits of a computer program that generates the output from the input) of the computer program that represents the Memory Unit. From Zurek's Principle we know that a computer program of a given *algorithmic complexity* can be represented by an ensemble of physical systems with entropy just equal to the *algorithmic complexity* of the computer program. We will design our physical Memory Unit so that it corresponds to the physical systems representing the computer program. An increase in *algorithmic complexity* consequently increases the complexity of the design of the physical Memory Unit without necessarily increasing the amount of heat flow to the bath. Therefore, excess and unnecessary information in the measurements leads to increased complexity in the design of the Memory Unit. A clean design thus requires that the measurements contain the minimum amount of information necessary to generate the output. The minimal information we need in a measurement is whether a particle approaching the trapdoor is on the correct side or not. The probability that the approaching particle is on the incorrect side is exactly equal to the probability that the trapdoor will open for the particle and allow it to pass. Therefore, the amount of information to minimally measure the particle is $I_t$. The amount of information flowing in the measurement communication channel that optimizes the design of the Memory Unit is exactly equal to the information flowing in the communication channel associated with the control signal.

## 5. Design of the Memory Unit

We imagine that the Memory Unit is an idealized physical device. This allows us to calculate the entropy of the device from ensemble considerations. From Zurek's Principle, we know that entropy calculated from this approach is equivalent, if optimal, to a deterministic calculation of the *algorithmic complexity* of the computation performed by the device. We design the Memory Unit so that it is able to hold the same information at each time as the control signal. This is an optimal design if the control signal is designed to hold only the minimum information needed to control the trapdoor. From Bennett's Principle we know that we can disregard for the entropy and energy accounting any computation that may be used to generate the output in the Memory Unit.

The details of the Memory Unit are displayed in FIG 2. The Memory Unit is a cylinder in a heat bath of temperature $T_B$. The cylinder contains a single particle. There are two pistons, one on each end of the cylinder. There are stops initially placed at the center of the cylinder. The pistons are allowed to compress to the stops. If a right piston is compressed and the left piston is not, then the control signal is ($o$), the trapdoor should be open. If the left piston is compressed and the right is not then the signal is ($c$), the trapdoor should be closed. The Unit is erased if both pistons are uncompressed. The left half of the Memory Unit is scored with N/2 evenly spaced ticks. New stops are placed one tick to the left after a piston compresses from the right. New stops are placed at the current tick when the piston compresses from the left. The old stops are removed when the Memory Unit is erased and the pistons return to their respective uncompressed states. When all the particles are sorted, the stops are at the left wall of the cylinder. The location of the stops, therefore, is measure of the degree to which the particles have been sorted.

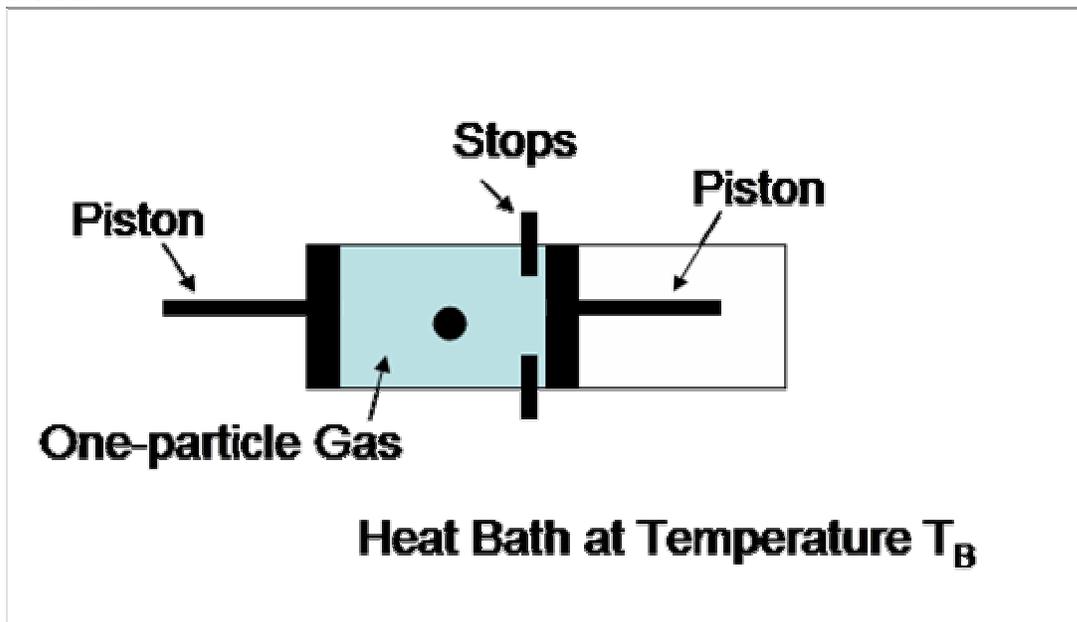

**FIG 2: The Memory Unit. The left half of the volume is divided into N/2 ticks (not shown). Stops are initially placed at a point halfway into the volume. Each time a particle is allowed to pass through the trapdoor, the stops are moved one tick to the left. This unit is indicating that an A type particle is approaching the trapdoor from the right (or a B type particle is approaching from the left) and that the trapdoor should open and allow passage. If an A particle approaches from the left (or a B particle from the right) the left piston is compressed against the stops and the right piston is against the right wall. This is the signal that the trapdoor should remain closed. A Unit that has been erased has both pistons at the edge of the containers.**

With this design for the Memory Unit, we can imagine a very simple mechanical communication channel for the control signal. This example can help clarify the differences between information communication and entropy generation. Assume that the pistons are connected to the trapdoor through a pulley system by two strings, one for each piston. The strings are held taut by finite but negligible masses. We assume the trapdoor and the strings are massless. The strings are connected in such a manner that when the right piston compresses, the trapdoor is pulled open by the string attached to the right piston. When the right piston is uncompressed, the string/spring attached to the left piston pulls the trapdoor to a closed position. There is no dissipation in the communication. Moreover, the state of the trapdoor (open or closed) is correlated with the state of the Memory Unit as it would be in the design of any reasonable communication channel. The entropy of the trapdoor is not calculated independently from the Memory Unit. Since the correlation is complete, a calculation of the entropy of the memory unit includes the entropy of the trapdoor. The same arguments can be made for the measurement communication channel. The signals in the communication channels, if dissapationless, do not generate entropy.

We can think of our string and mass system as a type of temperature coupling between the heat bath and the trap door. Small fast fluctuations of the pistons lead to small fast oscillations of the trapdoor. If the thermalization time of the pistons is small compared with the average time between particle encounters with the trapdoor, then the trapdoor is effectively at the same temperature as the heat bath. From Feynman's well-known ratchet and pawl arguments[10] we know that for the trapdoor to be effective it must be at a lower temperatures than the particles in the box. Therefore, the heat bath must be at a lower temperature than the particles in the box. Rather than a thermal coupling of computer to the trapdoor through the communication line, we can imagine that the trapdoor is directly coupled to the heat bath at a temperature lower than the temperature in the box.

We can now calculate the expected energy and entropy flows from writing to an output tape. If we imagine an ensemble of Memory Units, then the expected amount of work done by a piston, from a given side compressing a given volume, is

$$\Delta w_{\text{piston}} = k_B T_B \ln\left(\frac{V_u}{V_c}\right). \tag{5.1}$$

Here $k_B$ is Boltzmann's constant, $T_B$ is the temperature of the heat bath, $V_u$ is the volume of the uncompressed one-particle gas, and $V_c$ is the volume of the compressed gas. An equal amount of heat is transmitted to the thermal bath thus increasing the entropy of the bath by an amount

$$\Delta s_t = k_B \ln\left(\frac{V_u}{V_c}\right). \tag{5.2}$$

Note that if the piston is from the right and $\Pi_t$ is the fraction of particles on the wrong side of the box then the compression is

$$\Pi_t = \frac{V_c}{V_u}.$$

If the piston is from the left, then the compression is

$$1-\Pi_t.$$

the fraction of particles that are on the correct side of the box. Here $\Pi_t$ is also the probability that the control signal will be "open" at time $t$ and $1-\Pi_t$ is the probability that the control signal will be "closed." The total expected work done at time $t$ by the ensemble of Memory Units is therefore

$$\left\langle \Delta w_{\text{piston}} \right\rangle = k_B T_B \ln\left(2\right) I_t$$

$$I_t \equiv -\Pi_t \lg\left[\Pi_t\right] - \left(1-\Pi_t\right)\lg\left[\left(1-\Pi_t\right)\right]$$

(5.3)

Here, $I_t$ is the missing information, in bits, in the control signal before a measurement is made. Thus our Memory Unit is able to hold the same information as the control signal. Since particles approach the trapdoor randomly and with equal probability, this is the optimal representation of the missing information in the control signal. Neither the Memory Unit nor the control signal can store information more efficiently than the current design.

The expected entropy increase of the heat bath at time $t$ is

$$\left\langle \Delta s_t \right\rangle = k_B \ln\left(2\right) I_t \qquad \text{(measured in nats)}$$

$$= I_t \qquad\qquad \text{(measured in bits)}$$

(5.4)

The entropy increase of the heat bath is simply the missing information in the control signal.

We can, considering the entire control loop, explicitly calculate the expected fraction of particles on the wrong side of the box at time $t$ as a function of the expected fraction at time $t-1$. The time dependence of the expected value of the fraction of particles on the incorrect side of the box is given by

$$\Pi_t = \Pi_{t-1}\left(1-\frac{1}{N}\right)$$

$$= \frac{1}{2}\left(1-\frac{1}{N}\right)^t$$

(5.5)

where $N$ is the total number of particles in the box. In obtaining Eq. (5.5) we have used yet another interpretation of $\Pi_t$, that it is the expected number of particles to pass through the trapdoor at time $t$.

The state of the system after the Memory Unit has been written and the Control Signal has been transmitted is

     a. one of the two pistons is compressed decreasing the entropy of the cylinder from its erased state,

     b. the trapdoor is either open or closed depending on the state of the Memory Unit (which piston is compressed),

     c. the stops are at a certain tick depending on how many particles have been sorted, (We place a new set of stops at the incremented position if the right piston is compressed. Otherwise we will place the new stops at the current position. The erasure process will remove the old stops.)

     d. and the entropy of the heat bath has increased by an amount of work that the piston has done divided by the bath temperature.

Before the next measurement and control action can be made, the Memory Unit must be erased .

# 6. Erasure of the Memory Unit

As Landauer noted, erasure of a Memory Unit must be an action that is independent of the state of the unit. Erasure is a many-to-one mapping and is logically irreversible. This implies that the transition is also thermodynamically irreversible.

In our case, the Memory Unit can be in one of two states, depending on which piston is compressed. The erased state is one in which neither piston is compressed. A set of actions that erases the Unit independent of the initial state of the Unit is:

     a. Place a membrane at the stops.

     b. Apply a force to both pistons that pulls them out to their uncompressed state.

     c. Remove the membrane allowing particles to irreversibly expand to the entire cylinder.

     d. Remove the old stops.

This allows the entropy of the cylinder to naturally return to its value in the erased state without any heat transfer to or from the bath. The complete cycle of writing and erasing the tape generates an expected heat flow of

$$q_t = k_B T_B \ln(2) I_t \qquad (6.1)$$

to the bath, consistent with Landauer's Principle.

## 7. The Total Entropy Change

The total entropy change of the bath due to the repeated writing and erasure of the Memory Unit is given by

$$\Delta S_B = k_B \ln(2) \sum_{t=0}^{\infty} I_t$$

$$= (N + \eta N) k_B \ln(2) \quad \text{(measured in nats)} \qquad (7.1)$$

$$= (N + \eta N) \quad \text{(measured in bits)}$$

where in the limit of large N, $\eta$ reduces to Eq.(1.1) and (1.2). The entropy change of the Memory Unit itself is zero.

The entropy change of the sorted particles is

$$\Delta S_S = -N k_B \ln(2) \quad \text{(measured in nats)}$$
$$. \qquad (7.2)$$
$$= -N \quad \text{(measured in bits)}$$

Therefore, the minimum total entropy change of the heat bath, the Memory Unit, and the sorted particles is

$$\delta S = \eta N k_B \ln(2) \quad \text{(measured in nats)}$$
$$\qquad (7.3)$$
$$= \eta N \quad \text{(measured in bits)}$$

This is the minimum total entropy increase of the universe due to sorting $N$ randomly distributed particles into two bins by the Maxwell Demon process and is the main result of this paper. The physical interpretation of the excess entropy is clear. It is due simply to the fact that a particle may encounter the trapdoor more than once. On average, each particle encounters the trapdoor $1 + \eta$ times.

## 8. Concluding Remarks

We have addressed a slightly idealized version of Maxwell's original problem. We have attempted to organize earlier work in a manner that leads to an explicit calculation of the entropy increase of the universe due to sorting in the manner of a Maxwell Demon. In doing this we have emphasized the importance of addressing the complete control loop in the process. This leads to our main result, which simply stated, is that there is a minimum entropy generation of the universe due to sorting particles in a Maxwell Demon manner given by Eq. (1.1) and Eq. (7.3).

It is worthwhile to examine the subtle interactions among all the meanings of probability that led to the result Eq. (7.3). The probabilities change meaning depending on the context. For instance, the quantity $\Pi_t$ has taken on several different meanings depending on which part of the control loop is being addressed. The quantitative equivalence of all the meanings is the glue that binds the arguments.

In the System, for instance, $\Pi_t$ takes on two meanings. It is the fraction of particles that have not yet been sorted at time $t$. In other words, $\Pi_t$ is the probability that if one randomly picks a particle from anywhere in the box, that one will find a particle on the incorrect side. This meaning leads to a net change in entropy of $-N$ bits in the system as a consequence of the sorting. Also $\Pi_t$ is the expected number of particles to pass through the trapdoor at time $t$. The quantitative equivalence of the two meanings allows for the solution of the time dependence, Eq. (5.5).

In the measurement communication channel, $\Pi_t$ is the probability that a particle on the incorrect side of the box will approach the trapdoor. This defines

$$I_t \equiv -\Pi_t \lg\left[\Pi_t\right] - \left(1 - \Pi_t\right) \lg\left[\left(1 - \Pi_t\right)\right]$$

as the capacity of the communication channel.

In the Memory Unit, $\Pi_t$ is the level of compression of the right cylinder. This definition of $\Pi_t$ is part of the design of the Memory Unit. Also, $1 - \Pi_t$ is the compression of the left cylinder. This defines the amount of work done by the cylinders and the amount of heat transferred to the bath. The heat transfer defines the amount of entropy increase of the bath due to writing and erasure of the Memory Unit. The entropy increase is $I_t$, which is the same as the capacity of the measurement communication channel.

In the communication channel associated with the control signal, $\Pi_t$ is the probability that the control signal will be open ($o$). The capacity of the channel is once again $I_t$. This completes the loop.

The various personalities of the probability $\Pi_t$ have led to the generation and destruction of entropy, the generation of heat, and the transmission of information. To make this clear we collect here the total entropy and information accounting (in bits):

| | |
|---|---|
| Change in entropy of the particles in the box (order creation) | $-N$ |
| Total information transmitted in measurement channel | $(1+\eta)N$ |
| Entropy change of the Memory Unit | $0$ |
| Entropy change of the heat bath (heat generation) | $(1+\eta)N$ |
| Total information transmitted in control channel | $(1+\eta)N$ |
| Total entropy change of the universe (particles + bath) | $\eta N$ |

The Maxwell Demon problem is fundamentally a control problem. The Demon has a goal -- to organize particles. It actually provides clarity, then, to discuss the entropy accounting anthropomorphically or in an engineering sense. To restate results, the Demon's goal was to create order in a disorganized system. Therefore, the Demon generated an information flow that was greater than the desired entropy decrease of the disorganized system. An amount of heat associated with entropy equal to this information was transmitted to the environment. Because of this "payment" to the environment, the Demon was allowed to use the information to increase the order of the disorganized system.

Order can be generated locally if two systems such as the particle system and the Memory Unit are able to properly interact. One can easily imagine an idealized engine capable of useful work powered by a Memory Unit such as we have created here. This is the subject of a future paper.

Also in a future paper we will demonstrate that our result is quite general and can be applied to nonphysical processes in which a Maxwell Demon-like entity orders any type of random system. In particular, we apply the results to a stock market in which specialized traders increase the liquidity of the market while simultaneously bringing order to the market.